\documentclass[12pt,preprint2]{aastex}
\usepackage{natbib}

\begin{document}

\title{NGC 2362: a Template for Early Stellar Evolution
  \footnote{Based on observations collected at the European Southern
    Observatory, La Silla, under project 66.C-0119.} }

\author{A. Moitinho} 
\affil{Observatorio Astron\'omico Nacional, UNAM,
  Apdo. Postal 877, C.P. 22800, Ensenada B.C., M\'exico}
\email{andrem@astrosen.unam.mx}

\author{J. Alves}
\affil{European Southern Observatory, Karl Schwarzschild Stra\ss\ e 2,
D-85748 Garching, Germany}
\email{jalves@eso.org}

\author{N. Hu\'elamo} 
\affil{Max-Planck f\"ur Extraterrestrische Physik, Giessenbachstrasse
  1, D-85741 Garching, Germany}
\email{huelamo@mpe.mpg.de}
   
\and

\author{C.J. Lada} 
\affil{Harvard-Smithsonian Center for Astrophysics,Mail Stop 72,
  Cambridge, MA 02138}
\email{clada@cfa.harvard.edu}

\begin{abstract}
  We present UBVRI photometry for the young open cluster NGC 2362.
  From analysis of the appropriate color-color and color-magnitude
  diagrams we derive the fundamental parameters of the NGC 2362
  cluster to be: age $=$ 5$^{+1}_{-2}$ Myr, distance $=$ 1480 pc,
  $E(B-V)=$ 0.10 mag. The cluster age was independently determined for
  both high mass ($2.1 - 36$M$_{\sun}$) and low mass ($0.7 -
  1.2$M$_{\sun}$) stars with excellent agreement between the ages
  derived using post-main sequence \citep{Girardi2000iso} and pre-main
  sequence \citep{Baraffe1998} evolutionary tracks for the high and
  low mass stars respectively.  Analysis of this cluster's
  color-magnitude diagram reveals a well defined pre-main sequence
  (covering $\Delta$ V $\sim$ 9 magnitudes in $V$ and extending from
  early A stars to near the hydrogen burning limit) which makes this
  cluster an ideal laboratory for pre-main sequence evolution studies.

\end{abstract}


\section{Introduction}

The evolution of pre-main-sequence (PMS) stars -- young contracting
stars on their way to the main sequence -- is poorly understood
although significant progress have been made in recent years
\citep[see][for reviews]{Palla1999,Chabrier2000}. Observationally,
these stars are also among the hardest to study because they are
usually heavily extincted by their parental molecular cloud and are
often seen in projection against bright H{\footnotesize II} regions.
 
The ideal laboratory of pre-main-sequence stellar evolution would then
be the youngest possible galactic cluster free from dust extinction
and nebula contamination, at a distance that permits detection of
low-mass members to at least the hydrogen burning limit.  An
investigation of such an ideal cluster would provide robust
observables to test pre-main-sequence models (luminosity functions,
color-magnitude, and color-color diagrams), free from completeness
corrections and scatter due to variable extinction and nebula
emission. In turn, the age of this very young cluster would be well
constrained, as would be its star formation history, and its
underlying initial mass function (IMF).  The age of this cluster would
also set an upper limit on the timescale for the disruption of the
molecular material that originated the cluster.  Finally, observations
from X-rays down to the mid-Infrared wavelengths would also be able to
provide a transparent characterization of an entire pre-main-sequence
population down to brown dwarf masses, while shedding new light on the
X-ray/UV/H$_\alpha$/infrared-excess phenomena observed in young stars
\citep[see][for reviews]{Bertout1989,Feigelson1999}.

In this letter we present a UBVRI survey of the stellar cluster NGC
2362, probably the best proxy to the ideal cluster described above.
The cluster is virtually free from dust extinction and shows no signs
of nebular emission.  Walter Baade suggested more than fifty years ago
that NGC 2362 appeared to be almost exclusively made up of B stars
with little evidence for low mass stars \citep{Johnson1950}. This same
population of B stars has been used as the standard observational
template to define the upper end of the Zero Age Main Sequence.
Despite the claims for an abnormal mass function, a deep $I-$band
study of the central $6^\prime \times 6^\prime$ region of the cluster
\citep{Wilner1991} uncovered a substantial population of lower mass
stars whose mass function appeared to be similar to that of the field
\citep{Kroupa1992}.  Recently, \citet{Alves2001} presented sensitive
JHK observations of NGC 2362 that confirm this finding.  More
interesting perhaps, they found virtually no stars with detectable
near--infrared excess emission, suggesting that this young cluster
largely consists of diskless, post-T Tauri stars. In a survey of young
clusters at L band (3.4 $\mu$m), \citet{Haisch2001} confirmed this
result and further found NGC 2362 to be the youngest known cluster
without a significant population of disk bearing stars.  Consequently,
NGC 2362 plays a pivotal role in the determination of the overall
lifetime of circumstellar disks in clusters and for setting the
timescale allowed for planet building within such disks.
Unfortunately, the age of this cluster is not well constrained.  The
best existing age estimate essentially rests on the position in the HR
diagram of a single object, $\tau$CMa, the cluster's only post-main
sequence star \citep{Balona1996}. An improved age determination for
the cluster is urgently needed.

Our new deep UBVRI survey of NGC 2362 reveals a long, narrow, and well
defined pre-main-sequence, spanning about 9 magnitudes, in the $V$
band, from A-stars down to about $0.15$M$_{\sun}$, close to the
hydrogen burning limit. This enables us to provide an improved and
independent determination of the cluster's age. Comparison with modern
calculations of PMS evolution yields an age of $=$ 5$^{+1}_{-2}$ Myr
which closely agrees with the age we obtain from the cluster's early
type stars as derived from post-mais-sequence models. Our observations
of the early type stars also yield new estimates for the cluster's
distance and extinction which we find to be in good agreement with
previous determinations.

The remaining of this letter is organized as follows: we present the
observations in Section 2, the results in Section 3, and conclude with
a discussion on Section 4.

\section{Observations and Reductions}
\begin{figure}[t]
  \plotone{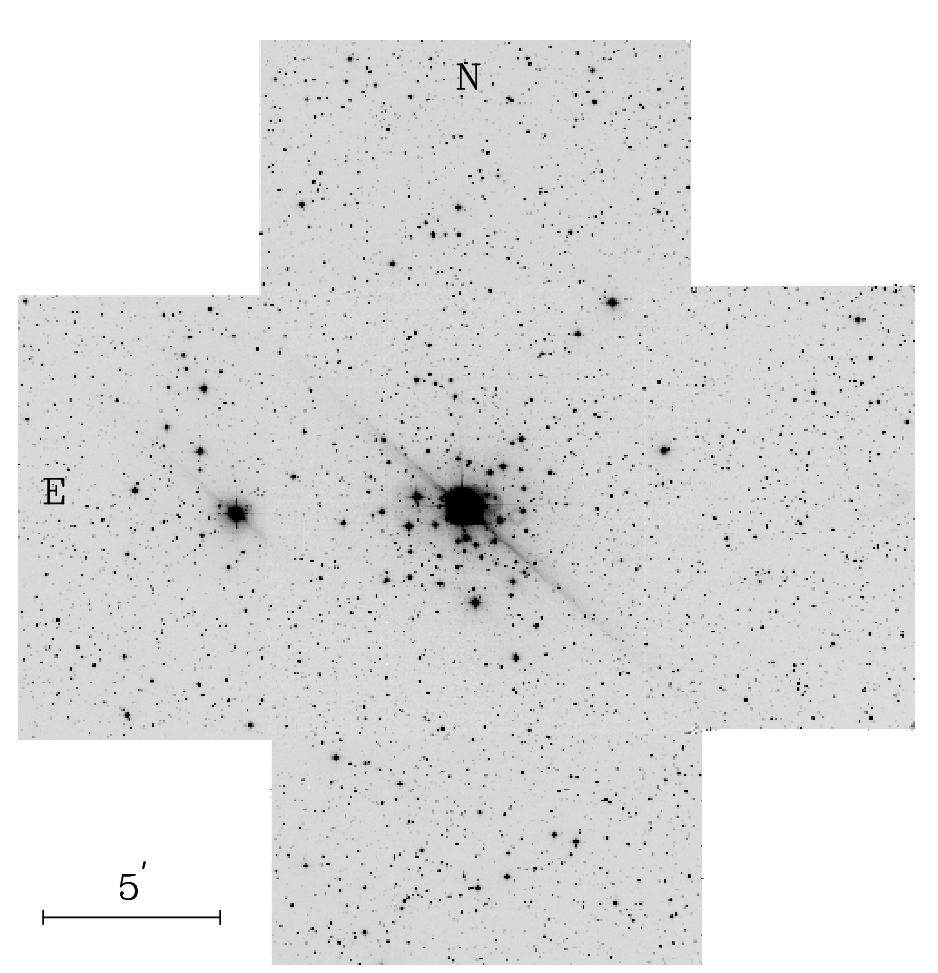}
  \caption{Mosaic of 15s exposures of NGC2362 in the V band. North is
    up and east is left. The surveyed area covers $\sim 540^{\prime 2}$
      The central $\sim 1.2^{\prime 2}$ portion of the mosaic, 
       which includes $\tau$CMa, is composed of a 1s exposure.}
\label{fig:mosaic}
\end{figure}
 CCD UBVR and Gunn {\it i} optical data were collected with DFOSC
mounted at the 1.5m Danish telescope in La Silla during the nights of
2001 February 2 and 4. The field-of-view of
DFOSC was $13.7^{\prime }\times 13.7^{\prime }$ with a plate-scale of
0.39$^{\prime\prime}$ per pixel. Due to severe distortions on the
edges, the images were trimmed which resulted in an effective
field-of-view of $12.9^{\prime }\times 13.3^{\prime }$.  A total of 36
images for the cluster, 10 images for the control field, and 36 images
of standard fields, were obtained.

To overcome the observational difficulties imposed by $\tau$CMa
(severe bleeding, scattered light, ghost images) resulting in
significant data corruption at the very center of the cluster
($\alpha_{2000} = 07^h18^m46^s.3$, $\delta_{2000} =
-24^{\circ}57^{\prime}22^{\prime\prime}$), a series of exposures of
NGC\,2362 were taken at four positions (NSEW) as close as possible to
$\tau$CMa but avoiding it or its scattered light (see
Fig.~\ref{fig:mosaic}).  During the same night we acquired short
exposures (1 sec) centered on $\tau$CMa to get photometry for the
bright stars at the center of the cluster and, in particular, those
that were in the region (an area of $\sim 1.2^{\prime 2}$) not covered
by the NSEW mosaic.

A nearby control field ($\sim 32^{\prime}$ north of the cluster) was
also observed, with the same exposure times as the cluster field
($280, 25, 15, 15$ and $40$s in the $U,B,V,R$ and Gunn {\it i} bands,
respectively), during the night of February 4th.  The control field
observations covered one field of view of the CCD.  Although the
integration times for the cluster and control field were the same, the
control field data is not as deep due to the proximity of the moon
during the control field observations.  Calibration into the standard
system was achieved by the observation of several \citet{Landolt1992}
fields. The typical seeing (FWHM) throughout the run was $\sim
1.1^{\prime\prime}$.

All CCD frames were processed within IRAF. Images were subjected to
the the usual bias and flatfield corrections, and cosmic rays were
removed. Photometry was performed with the IRAF/DAOPHOT package.  A
$7.8^{\prime\prime}$ ($20$ pixel) aperture radius was used in standard
star photometry.  The photometry, aperture corrections, extinction
corrections, transformation to the standard system and construction of
the photometric catalog were performed following the procedures
thoroughly described in \citet{Moitinho2001a}.  The final catalog
contains photometry for approximately 8300 stars ($\sim$ 4500 stars in
all UBVRI bands) in the cluster field, plus approximately 2000 stars
in the control field.

\section{Results}   
\begin{figure*}[htbp]
  \epsscale{1.4} 
  \plotone{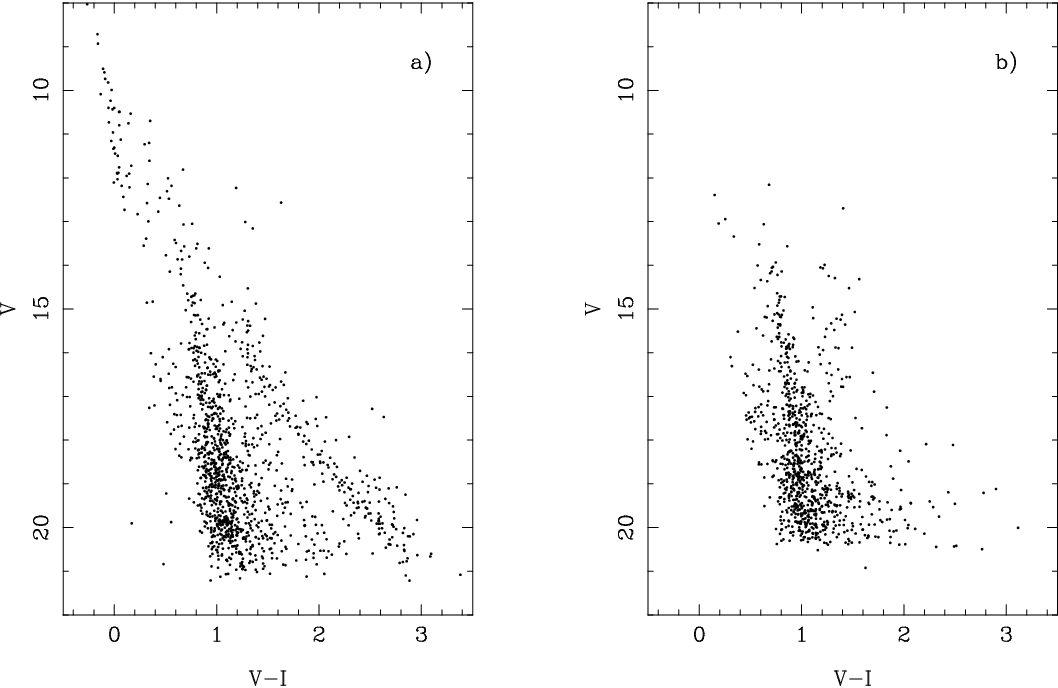}
  \caption{a) V vs. (V-I) diagram of the inner $9.1^{\prime}$ of the
    field of NGC\,2362. b) The same as in (a) for the control field.
    The control field data is not as deep due to the proximity
    of the moon during the control field observations.}
\label{fig:hr}
\end{figure*}
In Fig.~\ref{fig:hr}a we introduce the $V$ vs.  $(V-I)$ diagram of the
stars contained in a $9.1^{\prime}$ diameter circle centered on
NGC\,2362 (hereafter the \emph{on} field).  The center of the cluster
was estimated to be the peak of the stellar density distribution. The
distribution was computed using the stars falling along the cluster
sequence delineated in Fig.~\ref{fig:hr}a (both ZAMS and PMS stars).
The center is practically coincident with $\tau$CMa. Although previous
estimates of the cluster's diameter give a diameter of $5^{\prime}$
\citep{Lynga1987}, we find a significant concentration of stars, above
the background, in the inner $9.1^{\prime}$ diameter.

Fig.~\ref{fig:hr}b shows the $V$ vs. $(V-I)$ diagram for the control
field (hereafter the \emph{off} field). The area covered on the sky is
the same as in the left panel.  The distribution of stars in the
\emph{on} field color-magnitude diagram follows a sequence which
splits into two clearly separated branches below about $V$ = $15$
magnitudes.  The red branch of this sequence extends at least from $V
\sim 15$ to $21$ and has no counterpart in the \emph{off} field
color-magnitude diagram.  This red branch is part of a very long PMS
band of stars which starts at about $V \geq 12$ mag and is clearly
visible from $V \geq 15$ to $V \sim 21$ mag.

The reddening towards NGC\,2362 was determined by fitting the
\citet{Schmidt-Kaler1982} ZAMS to the B-star sequence in the $(U-B$) vs.
$(B-V)$ diagram. We have determined $E(B-V)=0.10 $ mag, which is in
agreement with the previous estimates by \cite{Perry1973} and
\cite{Balona1996}. No evidence of variable extinction was found 
at a $0.04$ mag level in $E(B-V)$.

The distance to NGC\,2362 was determined by fitting the
\citet{Schmidt-Kaler1982} ZAMS to the B-star sequence in different
color-magnitude diagrams. Unlike older clusters, where the A type
stars are already on the main sequence, the ZAMS stars in NGC\,2362
are almost exclusively of spectral type B. This makes the distance
determination more difficult since the B-star portion of the ZAMS is
nearly vertical for most color-magnitude combinations.  The color that
provides a B-star sequence with the lowest possible slope is $(U-B)$.
Therefore, distance was mainly constrained by the fit in the $V$ vs.
$(U-B)$ plane.  The apparent distance modulus was found to be
$(m-M)=11.16$ mag. Using the standard value for the ratio of
total-to-selective absorption, $R_V=3.1$, we find the distance to NGC
2362 to be $1480$ pc ($(m-M)_o=10.85$) which is virtually the same as the
value adopted by \citet{Balona1996} ($(m-M)_o=10.87)$. 

The age of NGC\,2362 was determined following two different
approaches.  In the first case, we have derived a turn-off age by
fitting solar composition post-main-sequence isochrones of
\citet{Girardi2000iso} to color-magnitude diagrams (using various
color-magnitude combinations).  In the second case, the
pre-main-sequence isochrones of \citet{Baraffe1998} (solar recipe)
were used in deriving an age based on the location of the low-mass PMS
stars in the $V$ vs. $V-I$ plane. These particular PMS models were
selected because they are the most consistent with constraints imposed
by recently derived dynamical masses of PMS binary stars
\citep{Steffen2001}. The solar recipe version of the PMS models has
only been computed for the $0.7 - 1.2$M$_{\sun}$ mass range, which was
the region used in estimating the age. The \citet{Baraffe1998} models
have also been computed for lower masses ($< 0.7$M$_{\sun}$) using non
solar mixing lengths (which is unimportant at such low masses).
However, the predicted optical colors at these cool temperatures are
not robust (indeed we found departures from the observed sequence),
and therefore have not been used in the age fits.

The youth of NGC\,2362 poses two problems to the turn-off
age determination. First, the observed B-star sequence does not show
any evident evolutionary deviation from the ZAMS (in fact, NGC\,2362
has been used in the past to define the upper ZAMS).  Second, the
high-mass end is dominated by poor statistics, which adds extra
difficulty to the analysis: The only star that exhibits evolutionary
effects is $\tau$CMa which, to make things even harder, is found to be
a multiple system \citep[eg.][]{vanLeeuwen1997}.

\begin{figure*}[htbp]
  \epsscale{1.4} 
  \plotone{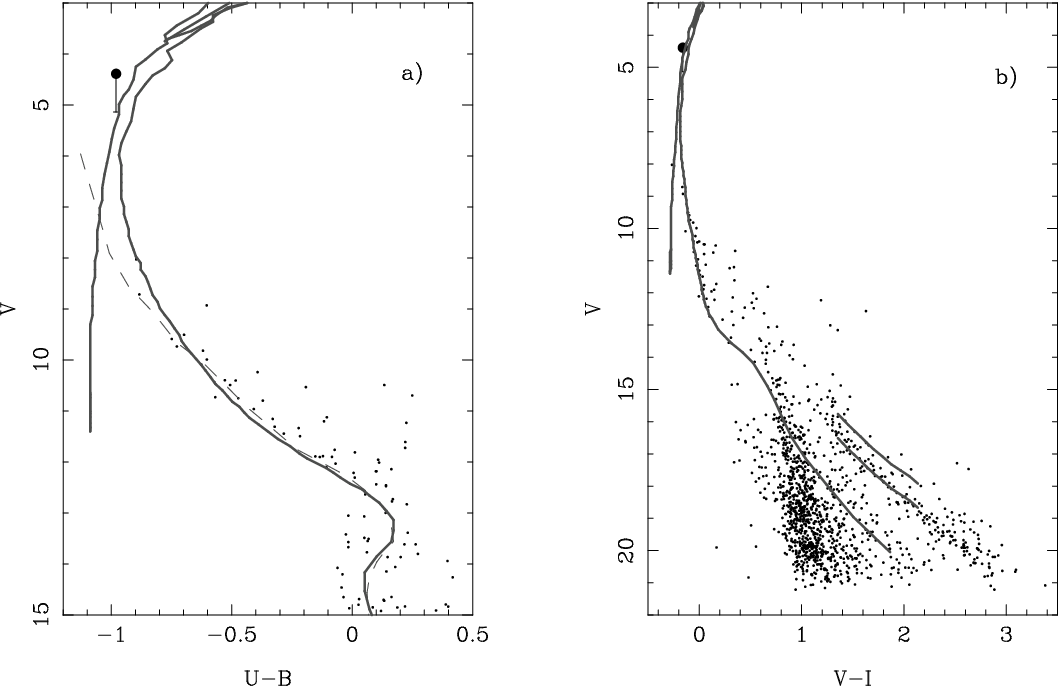}
  \caption{a) $V$ vs. $U-B$ diagram for the central $9.1^{\prime}$ of
    NGC\,2362. The filled line is a \citet{Girardi2000iso} 5 Myr post-
    main sequence isochrone. The dashed line is the
    \citet{Schmidt-Kaler1982} ZAMS used in the distance determination.
    b) isochrone fits in the $V$ vs. $V-I$ plane. The line on the left
    is the 5 Myr post-main sequence isochrones as in frame (a). The
    lines on the right are 5 Myr PMS isochrones \citep{Baraffe1998}.
    The brighter PMS isochrone has been plotted 0.75 mag brighter to
    account for the binary population. In both plots, the faint end of
    the bar appended to $\tau$CMa (large circle) indicates the
    magnitude of its equivalent single (non-multiple) star.}
\label{fig:age}
\end{figure*}
Fig.~\ref{fig:age} shows a 5 Myr isochrone fit in the $V$ vs. $U-B$
and $V$ vs. $V-I$ planes. The filled circle in the bright end is
$\tau$CMa, which has been plotted using the photometry of
\citet{Fernie1983}. A bar reaching 0.75 mag fainter indicates the
magnitude it would have if it was a simple (non-multiple) system.  The
fit has been essentially constrained by the position of $\tau$CMa
considering that it is a system of two similar - O9Ib - stars. This
age determination agrees with the one by \citet{Balona1996}, which was
also based on the position of $\tau$CMa.

As previously mentioned, we have also determined the cluster's age by
fitting pre-main sequence isochrones of \citet{Baraffe1998} to the
low-mass sequence on the $V$ vs. $V-I$ diagram (Fig.~\ref{fig:age}).
In this case, none of the problems of the turn-off fit arise: the
sequence is well populated, and the evolutionary status is quite
obvious (the PMS stars are well above the main sequence). Furthermore,
the luminosity of the PMS branch is a sensitive function of age.
Also, the absence of variable reddening within the cluster results in
an extremely well defined PMS which also helps to constrain the age.
The best fit was obtained for a 5 Myr isochrone (lines on the right of
Fig.~\ref{fig:age}). A 0.75 mag brighter PMS isochrone was also
plotted to account for the presence of binaries.  That the PMS is so
tight (well below the binary limit) is indicative of a very simple
star formation history, most probably a single and quick episode of
star formation. The age spread of the PMS stars in the cluster is
likely less than 3 Myr (see below).

To estimate the age uncertainty we have not considered the
uncertainties in the models themselves.  Taking into account the
photometric and fitting errors we estimate an uncertainty of
${}^{+1}_{-2}$ Myr.  The low dispersion of the PMS, together with the
sensitivity of the PMS luminosity to age, have allowed us to place tight
constraints on the age.  A somewhat higher uncertainty towards a
younger age arises from the fact that we fit the PMS isochrones to the
{\em lower/bluer envelope} of the observed sequence. Although the {\em
  lower/bluer envelope} fitting is usually performed in order to avoid
the effects of binary contamination and rotation, it could lead the
PMS fit towards slightly older ages if those effects are not
important.

We note here that the ages derived from the pre-main sequence and
post- main sequence isochrones are in excellent agreement with each
other as well as with the previous age determination for the cluster
which was based primarily on the position of $\tau$CMa in the HR
diagram (Balona and Laney 1996).  Moreover these age estimates are
also consistent with the dynamical age estimated for the
H{\footnotesize II} region surrounding the cluster and excited by
$\tau$CMa \citep{Haisch2001}.

\section{Conclusions}
We report UBVRI observations of NGC 2362, a very young open cluster in
CMa, and a nearby control field.  The main findings of this letter
are:
   \begin{enumerate} 
   \item We derive the following fundamental parameters for NGC 2362:
     age $=$ 5$^{+1}_{-2}$ Myr, distance $=$ 1480 pc, and reddening $=
     E(B-V)=$ 0.10 mag.  Cluster ages were obtained separately from
     both the high-mass population (OB-stars on the main sequence) and
     the low-mass (PMS) population of the cluster and found to be in
     excellent agreement.     
   \item Analysis of this cluster's color-magnitude diagram reveals a
     long, narrow, and well defined cluster pre-main-sequence,
     spanning about 9 magnitudes (in $V$) in the optical
     color-magnitude diagrams, with a negligible scatter.  This PMS
     extends from early A stars to late type stars near the hydrogen
     burning limit.
   \item Star formation in the cluster was likely characterized by a 
     single quick episode or burst spanning an interval of no more than
     about 3 million years.
   \item The observed properties of NGC\,2362 make it an ideal
     laboratory for testing pre-main sequence models and investigating
     early stellar evolution.   
   \end{enumerate}
   
   \acknowledgments We thank the referee for his/her useful comments
   which improved the paper.  A. Moitinho acknowledges financial
   support by CONACyT (Mexico; grant I33940-E) and DGAPA (Mexico;
   grant IN111500).  This research made use of the NASA Astrophysics
   Data System, of the Simbad database operated at the Centre de
   Donn\'es Stellaires --- Strasbourg, France, and of the WEBDA open
   cluster database of J.-C. Mermilliod.


\end{document}